\documentclass[twocolumn]{aastex61}
\usepackage{graphicx}
\usepackage{amsmath,amstext}
\usepackage[T1]{fontenc}
\usepackage{apjfonts} 
\usepackage[figure,figure*]{hypcap}

\shorttitle{Peak spectra of SLSN-I}
\shortauthors{Gal-Yam}

\begin{document}

\title{A simple analysis of Type I superluminous supernova peak spectra: composition, 
expansion velocities, and dynamics}

\author{Avishay Gal-Yam}
\affiliation{Department of Particle Physics and Astrophysics, Weizmann Institute of Science}

\begin{abstract}

We present a simple and well defined prescription to compare absorption lines in supernova (SN) spectra with lists of transitions drawn from the National Institute of Standards and Technology (NIST) database.
The method is designed to be applicable to simple spectra where the photosphere can be mostly 
described by absorptions from single transitions with a single photospheric velocity. These conditions 
are plausible for SN spectra obtained shortly after explosion. Here we show that the method also works well for spectra of hydrogen-poor (Type I) superluminous supernovae (SLSNe-I) around peak. Analysis of high signal to noise spectra leads to clear identification of numerous spectroscopic features arising from ions of carbon and oxygen, that account for the majority of absorption features detected in the optical range, suggesting the outer envelope of SLSN-I progenitors is dominated by these elements. We find that the prominent
absorption features seen in the blue are dominated by numerous lines of OII, as previously suggested, and that the apparent absorption feature widths are dominated by line density and not by doppler broadening. In fact, we find that while the expansion velocities of SLSNe-I around peak are similar
to those of normal SNe, the apparent velocity distribution (manifest as the width of single transition
features) is much lower ($\sim1500$\,km\,s$^{-1}$) indicating emission from a very narrow photosphere in velocity space that is nevertheless expanding rapidly. We inspect the controversial case of ASASSN-15lh, and find that the early spectrum of this object is not consistent with those 
of SLSNe-I. We also show that SLSNe that initially lack hydrogen features but develop these at late phases, such as iPTF15esb and iPTF16bad, also differ in their early spectra from standard SLSNe-I.  

\end{abstract}


\section{Introduction}

Supernovae (SNe) are identified and classified mostly using peak-magnitude optical spectra (\citealt{1997ARA&A..35..309F}; see \citet{2017hsn..book..195G} for an updated review). Spectroscopic classification relies on identification of specific spectral lines, for example, regular Type II SNe show strong hydrogen absorption lines, peak spectra of Type Ib SNe show strong He I absorption, and so on. Recently, automated template-matching routines (e.g., \citealt{2007ApJ...666.1024B}\footnote{SNID, \url{https://people.lam.fr/blondin.stephane/software/snid/}}; \citealt{2005ApJ...634.1190H}\footnote{Superfit, \url{http://www.dahowell.com/superfit.html}} and \citealt{2008A&A...488..383H}\footnote{Gelato, \url{https://gelato.tng.iac.es/}}) have become widely used. Such routines classify new objects based on their spectral similarities to known templates, but of course, the classification of the templates still relies on specific line identifications. Line identification is in particular essential when known types of SNe are observed for the first time in phases not covered by template libraries (such as ``flash spectroscopy'' obtained shortly after explosion, e.g., \citealt{2014Natur.509..471G}, \citealt{2007ApJ...666.1093Q} and \citealt{2017NatPh..13..510Y}), or when new types of events are encountered. 

A case in point are Superluminous SNe (SLSNe; see \citealt{2018arXiv181201428G} for a review). When the first examples of hydrogen-poor (Type I) SLSNe were discovered (\citealt{2007ApJ...668L..99Q}, \citealt{2009Natur.462..624G}, \citealt{2009ApJ...690.1358B}), the lack of known templates to compare with initially precluded classifiication of some events. Only the accumulation of several examples spanning a broad redshift range allowed \citet{2011Natur.474..487Q} to define this new class of events.

Historically, line identification in SN spectra is done manually; this is complicated by the need to account for the SN expansion velocity, and by confusion due to blending of lines of different elements that have very similar wavelengths. Sometimes the identifications are trivial and robust (e.g., identifying the hydrogen Balmer series based on several strong lines in the optical range) while in other cases it is more challenging (for example, the ongoing debate about the nature of the absorption feature observed near $\lambda6150$\AA\ in Type Ib and Ic events, e.g., \citealt{1997ARA&A..35..309F}, \citealt{2016ApJ...827...90L}, \citealt{2016ApJ...820...75P} and \citealt{2017hsn..book..195G}). Here, we present a method based on a simple and well defined prescription to compare absorption lines in supernova spectra with lists of transitions drawn from the National Institute of Standards and Technology (NIST) database. The method was developed toward
an application to spectra of very young SN explosions, that are expected to, and present, few features at a narrow range of photospheric expansion velocities, as recently demonstrated by \citet{2019arXiv190411009H}. Here we show that it works well for peak spectra of SLSNe-I, and discuss the results of this analysis.  

\section{A brief description of the method}

The goal of the method we present here is to identify the main spectral features in simple SN spectra. We define a spectrum as simple in this context as a spectrum that can be adequately described
by distinct spectral features that share a single expansion velocity. This means of course that spectral lines are not heavily blended together, and that the range of photospheric velocities of different ions is narrow (i.e., the photosphere is not deep in velocity space). As we show below, these conditions are met in some SN spectra, but are certainly not met in many other cases. In that respect our approach cannot replace detailed spectral synthesis modelling of complex spectra (e.g., \citealt{1993A&A...279..447M}; \citealt{2005A&A...437..667D}, \citealt{2000ApJ...545..444B}, \citealt{2011PASP..123..237T}). 

Our goal is to provide a robust, quick, and objective way to analyze simple spectra. In particular, in order to assess whether a certain ion contributes to an observed spectrum, we wish to create lists of strong line features that are objective and robust, in the sense that the choice of lines associated with each ion is well defined and reproducible, and that only strong lines are included in the list. This is helpful since in this case the lack of certain expected features can be used to argue against a contribution of a certain ion.  

We construct our line lists in the following manner. First, for each ion of interest we extract the list of lines from the National Institute of Standards and Technology (NIST) Atomic Spectra Database (ASD; \citealt{NIST-ASD-REF}\footnote{accessed through the following webform: \url{http://physics.nist.gov/PhysRefData/ASD/lines_form.html}}). We order the line list by the ``relative intensity'' parameter provided by the database, determine the maximum of this intensity parameter in the wavelength range of interest (for the current work we use $180-1000$\,nm), and include in our list lines with intensity values that are above some fraction (typically $50\%$) of the maximum value. Of course, the choice of this threshold is arbitrary, but we find it works well for our needs, and can be easily modified as needed. In a companion paper (Gal-Yam, Yaron \& Knezevic, in preparation), we describe in more detail our survey of other methods to compile line lists from the NIST databases and other tests and applications of this method. 

With the line lists in hand, we attempt to determine for each spectrum the expansion velocity by matching features to ions with numerous transitions (see. e.g., Fig.~\ref{Fig12damred}). Once we have determined a convincing value for the expansion velocity, we inspect other ions using the fixed velocity (i.e., we assume explicitly that the photosphere is narrow). 

\subsection{Application to peak spectra of SLSNe-I}

We begin our examination of SLSN-I peak spectra by inspecting high signal to noise (SNR) spectra of nearby events. In particular, we begin by focussing on the red side of the optical range 
($ \lambda > 5000$ \AA). This wavelength range has been relatively neglected in studies of SLSN-I spectra since it does not contain strong features, and is redshifted out of visible light spectra of SLSNe at moderate or high redshifts, which are typical for SLSNe. 

Figure 1 (left) shows our analysis results for a high SNR spectrum of PTF12dam from \cite{2013Natur.502..346N} obtained on 2012 May 24, 25 restframe days prior to peak according to the analysis of \cite{2017ApJ...835...58V}. This and all other spectra discussed here are available from WISeREP \citep{2012PASP..124..668Y}\footnote{\url{http://wiserep.weizmann.ac.il/}}. Multiple features consistent with OI and CII lines all align using a single expansion velocity of V$_{\rm exp}=11000$\,km\,s$^{-1}$; these two ions explain most features in this wavelength range. CIII may explain two weak features, and may also contribute to the observed spectrum. OII has a distinct line spectrum with a multitude of lines all with very similar intensities. While a few single lines seen around $6500$\AA\ appear to be negligible compared to the stronger CII lines, accumulations of multiple overlapping OII lines begin to shape the spectrum bluewards of $5000$\AA; we will return to discuss this ion promptly below. 

\begin{figure*}
\hspace*{-1.5cm}\includegraphics[width=10.5cm]{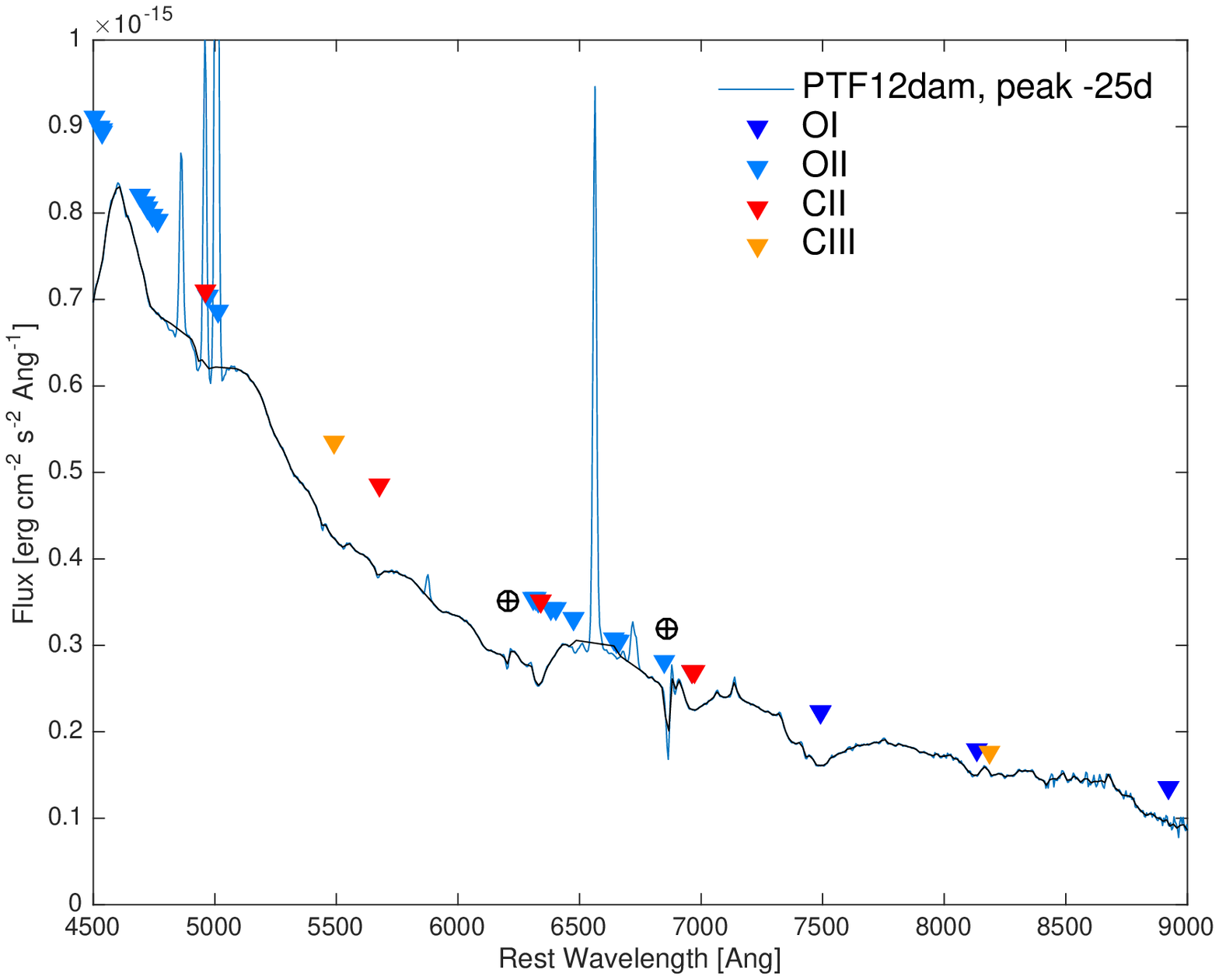}
\hspace*{-0.5cm}\includegraphics[width=10.5cm]{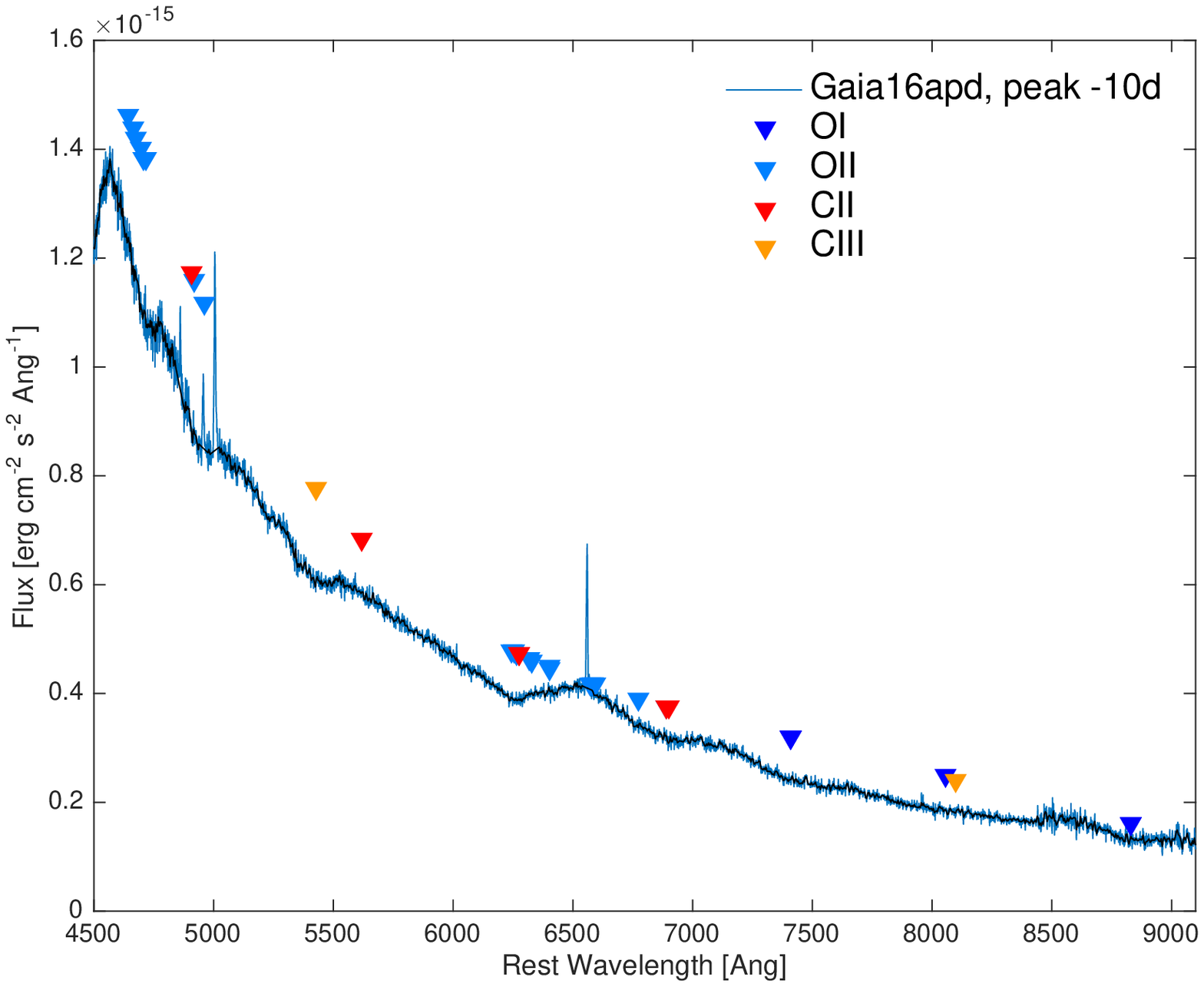}
\caption{High SNR red spectra of the nearby SLSN-I PTF12dam (left; from \citealt{2013Natur.502..346N}) and Gaia16apd (right; from \citealt{2017ApJ...840...57Y}). Both spectra (black curves) have been binned and host narrow emission lines have been excised; the original spectra are shown in blue. Uncorrected Telluric features in the spectrum of PTF12dam are marked. The spectrum of PTF12dam shows pronounced lines of OI and CII with distinct sharp absorption troughs, all at the same expansion velocity of V$_{\rm exp}=11000$\,km\,s$^{-1}$; OII lines probably begin to shape the spectrum blueward of $5000$\AA\, while CIII may also contribute. The same is true for Gaia16apd (right) though absorption lines are of lower contrast and have higher expansion velocity, V$_{\rm exp}=14000$\,km\,s$^{-1}$.}
\label{Fig12damred}
\end{figure*}

A similar result is obtained by analysis of the early spectrum (obtained on May 31, 2016, i.e., 10 days prior to g-band maximum according to \citealt{2017MNRAS.469.1246K}) of Gaia16apd (SN 2016eay; right panel of Fig.\ref{Fig12damred}), from \cite{2017ApJ...840...57Y}, where a preliminary application of our method is reported. Again, the same ions of C and O explain well almost all features with a single expansion velocity (V$_{\rm exp}=14000$\,km\,s$^{-1}$). Line features are shallower and have lower contrast, but are still easily visible in this high SNR spectrum. As noted by \cite{2018ApJ...853...57B}, their spectrum of the recent, nearby SLSN-I SN 2017egm, also shows the CII lines discussed here, where this identification is confirmed using modelling with SYNOW. Even stronger CII lines are seen in the spectra of the nearby SN2018bsz by \citet{2018A&A...620A..67A}. We conclude that analysis of the red portion of visible-light spectra of SLSN-I prior or around peak using this method gives consistent results, with the spectrum explained by C and O lines with a single photospheric velocity. The ability to determine the absolute expansion velocity (rather a velocity relative to template events) is a valuable result of this analysis. 

\begin{figure*}
\hspace*{-1.5cm}\includegraphics[width=10.5cm]{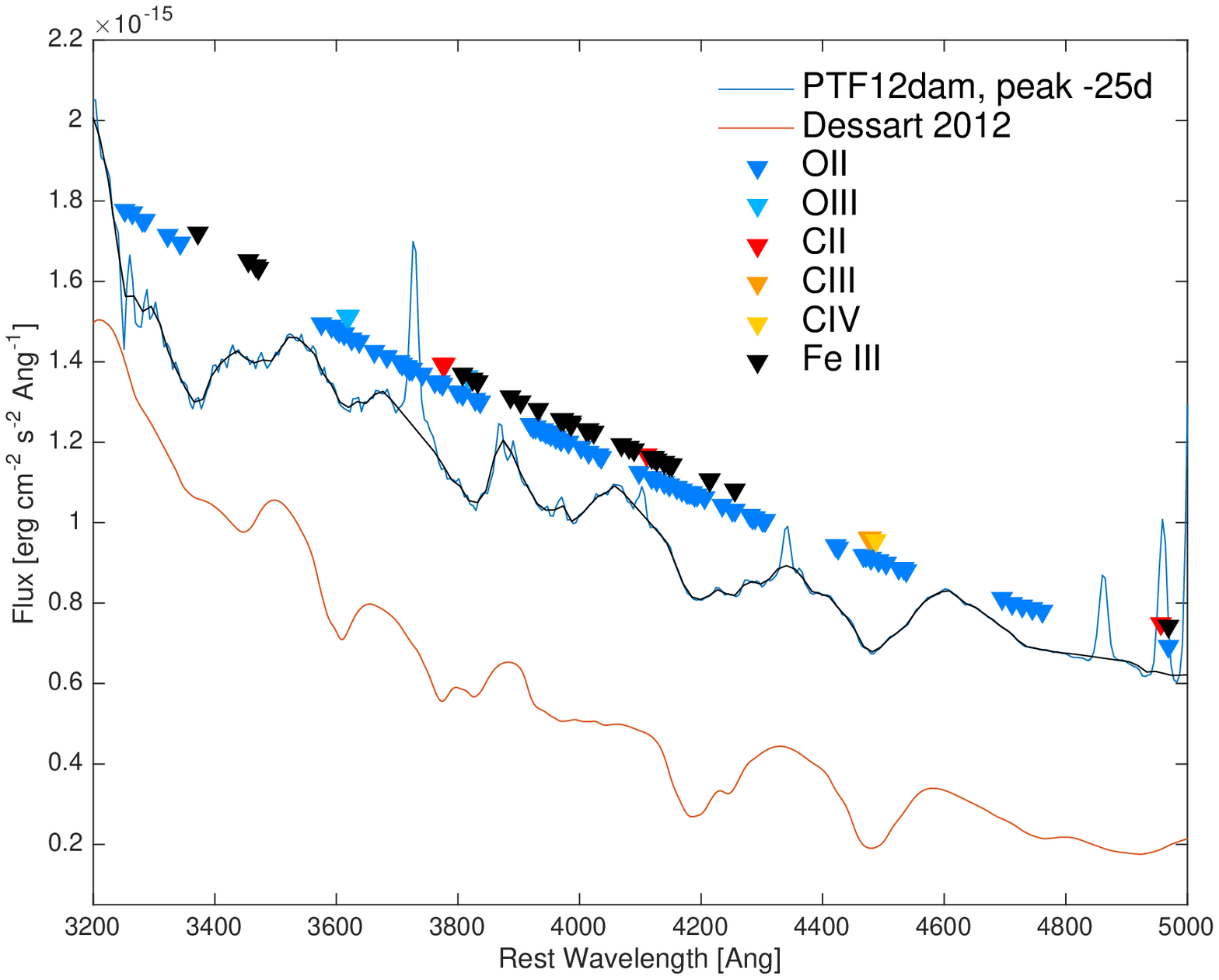}
\hspace*{-0.5cm}\includegraphics[width=10.5cm]{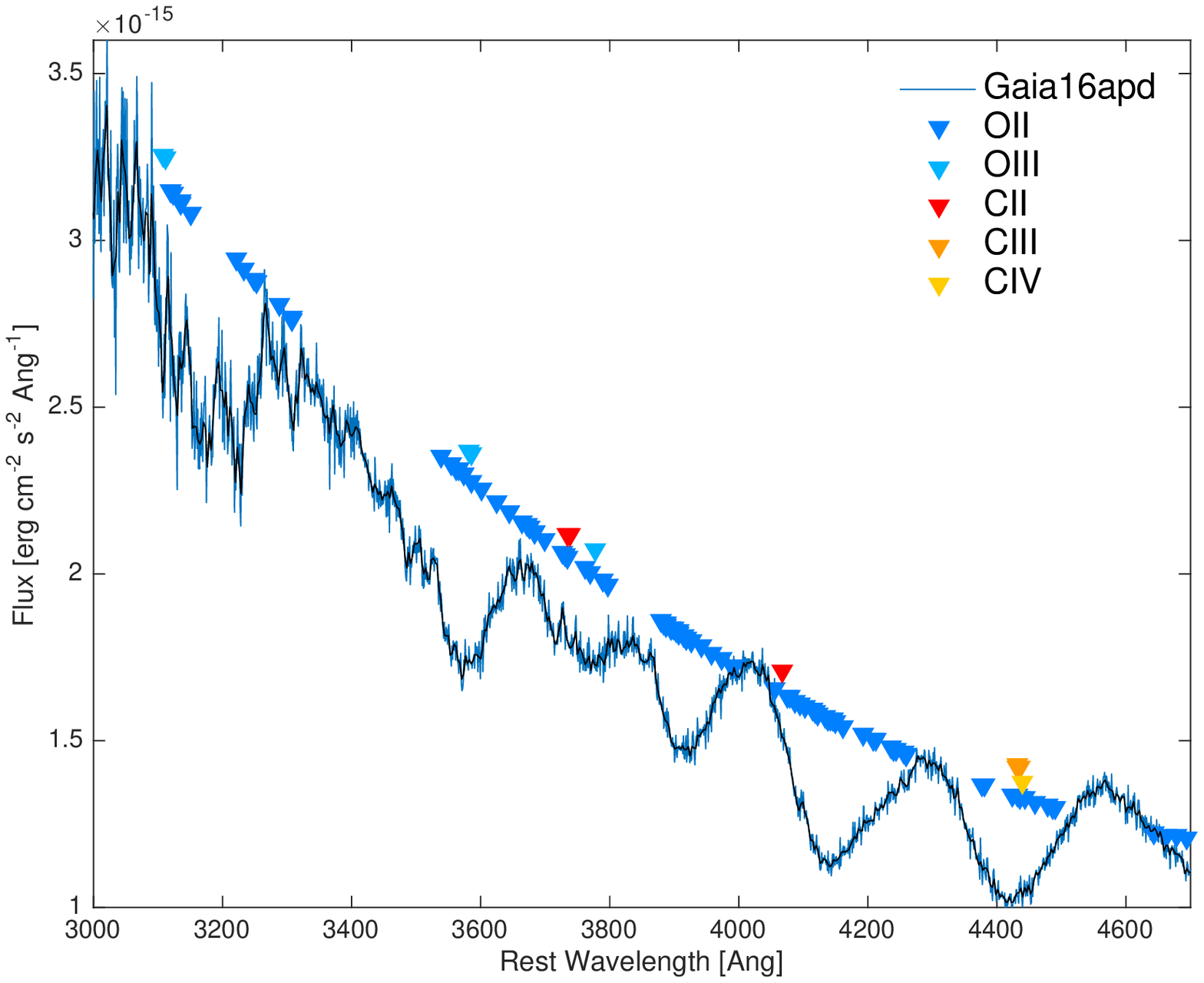}
\caption{High SNR blue spectra of the nearby SLSN-I PTF12dam (left; from \citealt{2013Natur.502..346N}) and Gaia16apd (right; from \citealt{2017ApJ...840...57Y}). Both spectra (black curves) have been binned and host narrow emission lines have been excised; the original spectra are shown in blue. 
The blue spectra of both PTF12dam and Gaia16apd are explained very well by a thick blanket of OII lines; with the apparent strong spectral features actually resulting from gaps in the OII line distribution manifesting as emission ``peaks''. Other C and O ions (marked) may also contribute, in particular OIII around $3600$\AA\ . The velocity deduced from single transitions in the red side (Fig.~\ref{Fig12damred}) aligns the OII ``gaps'' perfectly with the apparent spectral peaks, supporting our spectral line identifications. 
A synthetic model from \cite{2012MNRAS.426L..76D} is shown in the left panel for comparison; the line identifications based on this detailed model agree with the simple analysis presented here. Based on this model, the absorption between $3350-3600$\AA\ that is not accounted for by OII is due to Fe III (black). }
\label{Fig12damblue}
\end{figure*}

We now turn our attention to the blue portions ($\lambda<5000$\AA) of the same spectra, which are presented in Fig.~\ref{Fig12damblue}. Here, the spectrum seems to be formed by a blue continuum seen through a thick forest of OII absorption lines. The multitude of lines in the blue part of the visible light spectrum are all of similar intensity, and the shape of the spectrum is a function of line density rather than line strength; gaps in the OII line spectrum appear as emission peaks, while regions with dense line clusters appear as smooth valleys. Interestingly, the near ultra-violet (UV) portion of the spectrum which is well-sampled in the Gaia16apd spectrum (right) has a lower density of OII lines, leading to a ``saw-tooth'' structure replacing the smoother features in the redder parts. In particular, several features around $3300$\AA\ appear to result from single transitions, which are much narrower than the broader absorption valleys in the red. In the left panel of Fig.~\ref{Fig12damblue} we also overplot a synthetic model spectrum from \cite{2012MNRAS.426L..76D}; the line identification reported there, as well as in \cite{2016MNRAS.458.3455M}, agrees well with our line identifications, validating our simple approach. Based on this model, the absorption between $3350-3600$\AA\ that is not accounted for by OII is due to Fe III (black) which may also contribute to some of the redder features to some degree. 

\section{Results}

\subsection{Ejecta composition}

The identification of almost all major features in the visible light spectrum of SLSNe-I before or near peak with C and O lines (only) suggests that the outer parts of these powerful explosions are almost pure C/O envelopes. We do not detect any traces of neither hydrogen nor helium. Interestingly, also signatures of intermediate-mass elements (IME), such as magnesium, silicon and calcium, that are very prominent in spectra of normal SNe Ib and Ic (e.g., \citealt{1997ARA&A..35..309F}, \citealt{2017hsn..book..195G}, \citealt{2017MNRAS.469.2498M} and references therein), are not observed. The weak signatures of Fe lines are probably consistent with the natal metallicity of the progenitor \citep{2016MNRAS.458.3455M}, and the absence of  IMEs and iron-group elements (IGEs) many weeks after the explosion of these events suggests that nucleosynthetic products have not been mixed out into the outer envelope, and that the photosphere has not penetrated into deeper, enriched layers. A transition in the spectral appearance of SLSNe-I which is consistent with such inward movement of the photosphere has been observed in later phases (e.g., by \citealt{2010ApJ...724L..16P}, \citealt{2011Natur.474..487Q}, \citealt{2016ApJ...826...39N} and \citealt{2017ApJ...837L..14L}). Overall, our results confirm and reinforce the observations of \cite{2011Natur.474..487Q} that indicate that the initial radiation from SLSNe-I originates from almost pure C/O envelopes that extend to large radii ($>10^{15}$\,cm) and that this emission mode persists for many weeks.   
   
\subsection{Velocities}
\label{secvel}

\begin{figure}
\hspace*{-1cm}\includegraphics[width=10.5cm]{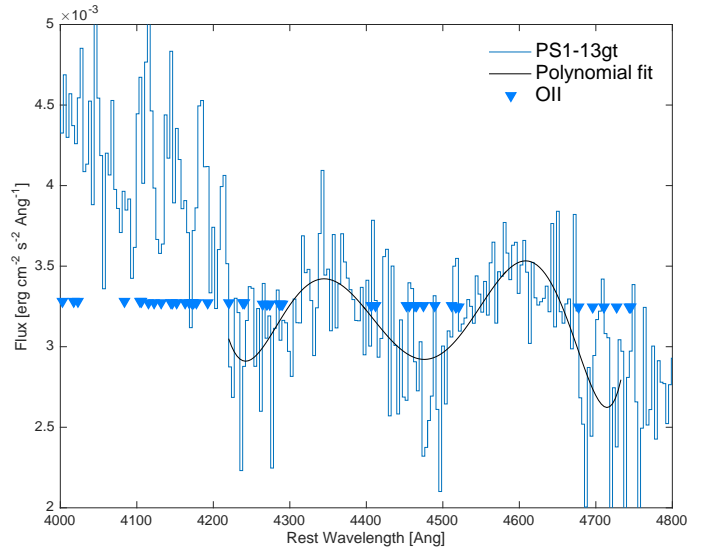}
\caption{Application of the method to the noisy spectrum of the high-redshift SLSN-I PS1-13gt. At z=0.884, this event is the most distant in the sample of \citet{2018ApJ...852...81L} showing the OII complex. Shifted into the red part of the visible light spectrum most affected by sky lines, the OII spectrum presents a challenge to analyse. Here we plot the original spectrum (dereddened by E$_{\rm B-V}=0.3$\.mag following the estimate of \citet{2018ApJ...852...81L} and binned for clarity; blue) and a high-order polynomial fit to the strongest two OII features (black) which excludes the rnage below restframe 4200\AA\ which is affected by a cluster of sky lines. Overplotting our OII line list we can estimate the expansion velocity of this object to be V$_{\rm exp}=12000\pm1000$\,km\,s$^{-1}$, with values ourside this range leading to mismatch between the line gaps and the observed emission peaks. It would seem difficult to apply this method on data of substantially inferior quality.}
\label{Fig_PS1_13gt}
\end{figure}

Our analysis allows us to determine the absolute expansion velocities of SLSN-I ejecta from the strongest spectral features in visible light spectra, the OII absorptions in the blue region, by aligning the gaps in the OII line distribution with the spectral peaks in this region (Fig.~\ref{Fig12damblue}). The absolute scale is established and validated by weaker, but definite, association of OI and CII lines in the red with individual transitions (Fig.~\ref{Fig12damred}). In the examples above, we find velocities of $11000$\,km\,s$^{-1}$ and $14000$\,km\,s$^{-1}$ for the spectra of PTF12dam and Gaia16apd at $-25$\,d and $-10$\,d before peak, respectively. Our method can be used to trace the ejecta velocity evolution of SLSNe-I based on a series of spectra, even if these are of relatively low SNR, using the strong OII features, as demonstrated in Fig.~\ref{Fig_PS1_13gt}. The typical velocity accuracy of this method, which we estimate by the minimal velocity shift that leads to a significant overlap of listed absorption lines over observed emission peaks, is $1000$\,km\,s$^{-1}$. Using spectra with well-determined velocities from narrow features (e.g., CII in Fig.~\ref{Fig12damred}) as templates for comparison with lower SNR spectra improves the velocity accuracy to $\approx500$\,km\,s$^{-1}$.  

\begin{figure}
\hspace*{-1cm}\includegraphics[width=10.5cm]{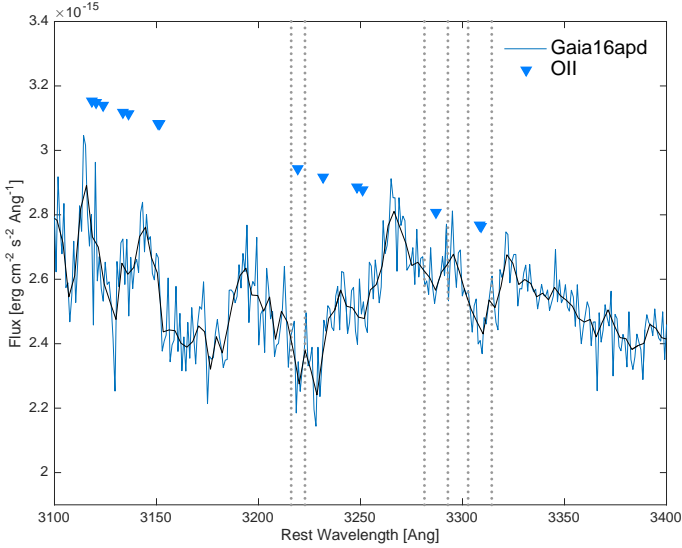}
\caption{A zoomed-in view of the blue part of the Gaia16apd peak spectrum showing absorption features from individual OII transitions. The narrow features indicate very low values of velocity dispersion - ranges of $\pm300$\,km\,s$^{-1}$ (left feature) and $\pm500$\,km\,s$^{-1}$ (right fatures) are marked.}
\label{Fig_gaia16apd_blue_zoom}
\end{figure}

\begin{figure}
\hspace*{-1cm}\includegraphics[width=10.5cm]{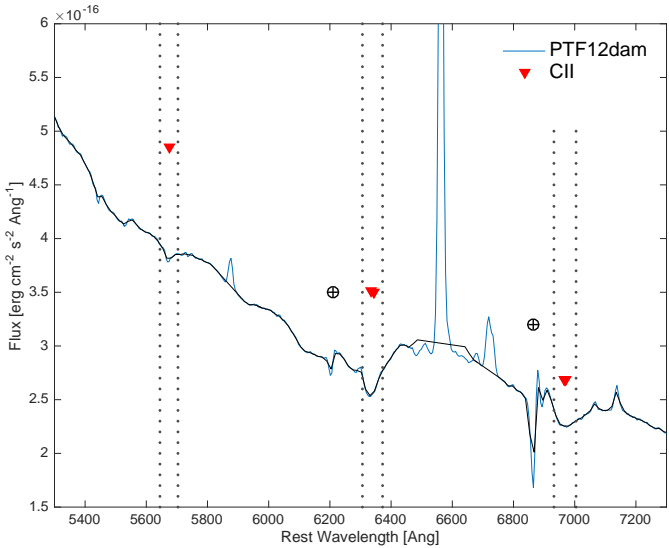}
\caption{A zoomed-in view of the red part of the PTF12dam -25\,d spectrum showing CII absorption features with narrow velocity dispersions; ranges of $\pm1500$\,km\,s$^{-1}$ are marked.}
\label{Fig_PTF12dam_red_zoom}
\end{figure}

As shown in Fig.~\ref{Fig12damblue}, the strong OII line feature shape is dominated by line density rather than ejecta velocity dispersion (as is usually the case for SNe). Toward the blue side of the spectra, the OII line density drops and some features can be identified with individual features (Fig.~\ref{Fig_gaia16apd_blue_zoom}). Interestingly, the derived line widths suggest a very low velocity dispersion, which is also consistent with the narrow cores of the weaker CII lines in the red (Fig.~\ref{Fig_PTF12dam_red_zoom}). This justifies our assumption that the peak spectra of SLSNe-I are simple, i.e., they have very shallow photospheres that can be described by a single velocity. \cite{2018ApJ...855....2Q} reach similar conclusions about the velocity widths of individual OII lines using a different and independent analysis. 

\subsection{Identification of the UV lines}

\begin{figure*}
\hspace*{-1.5cm}\includegraphics[width=10.5cm]{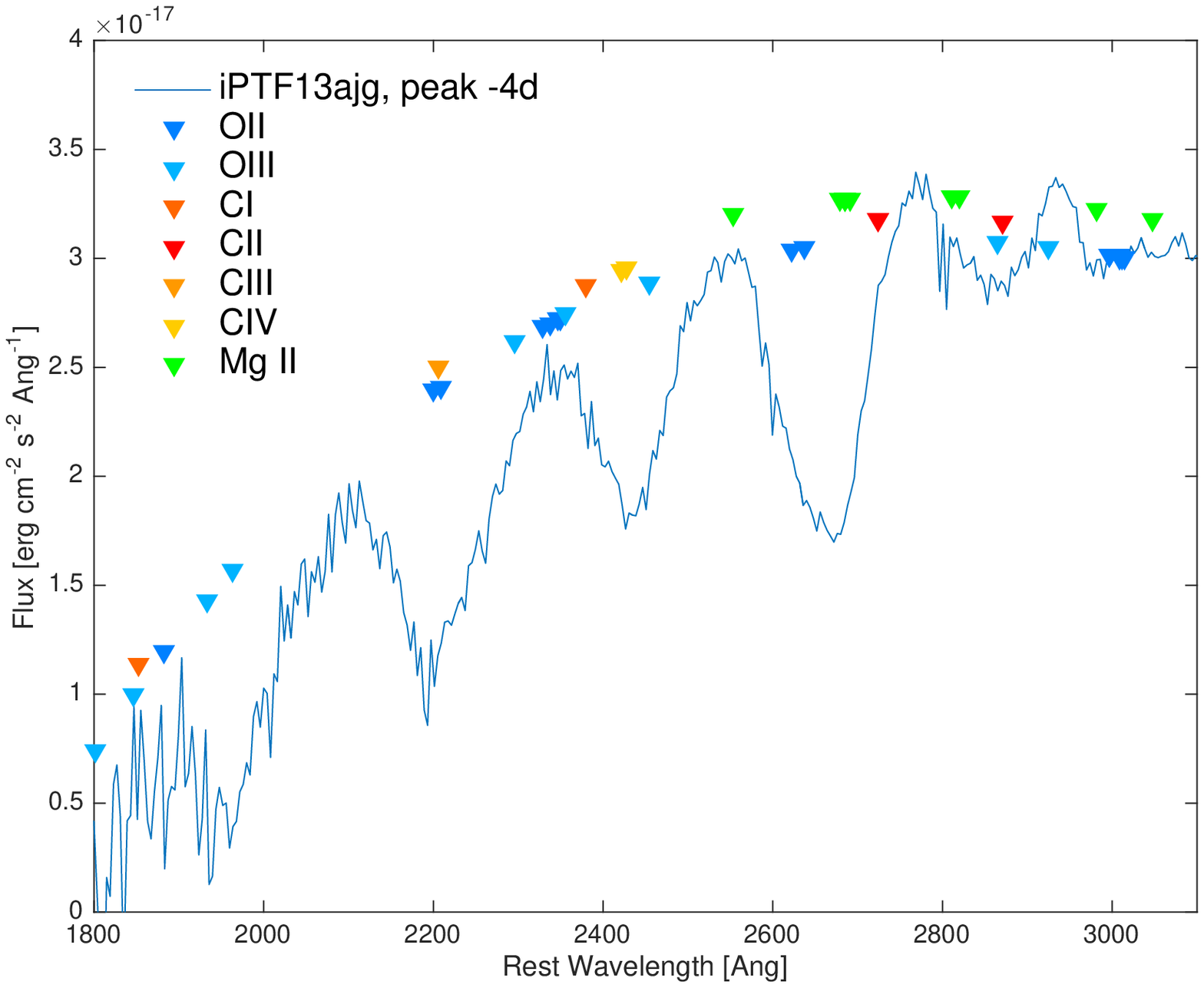}
\hspace*{-0.5cm}\includegraphics[width=10.5cm]{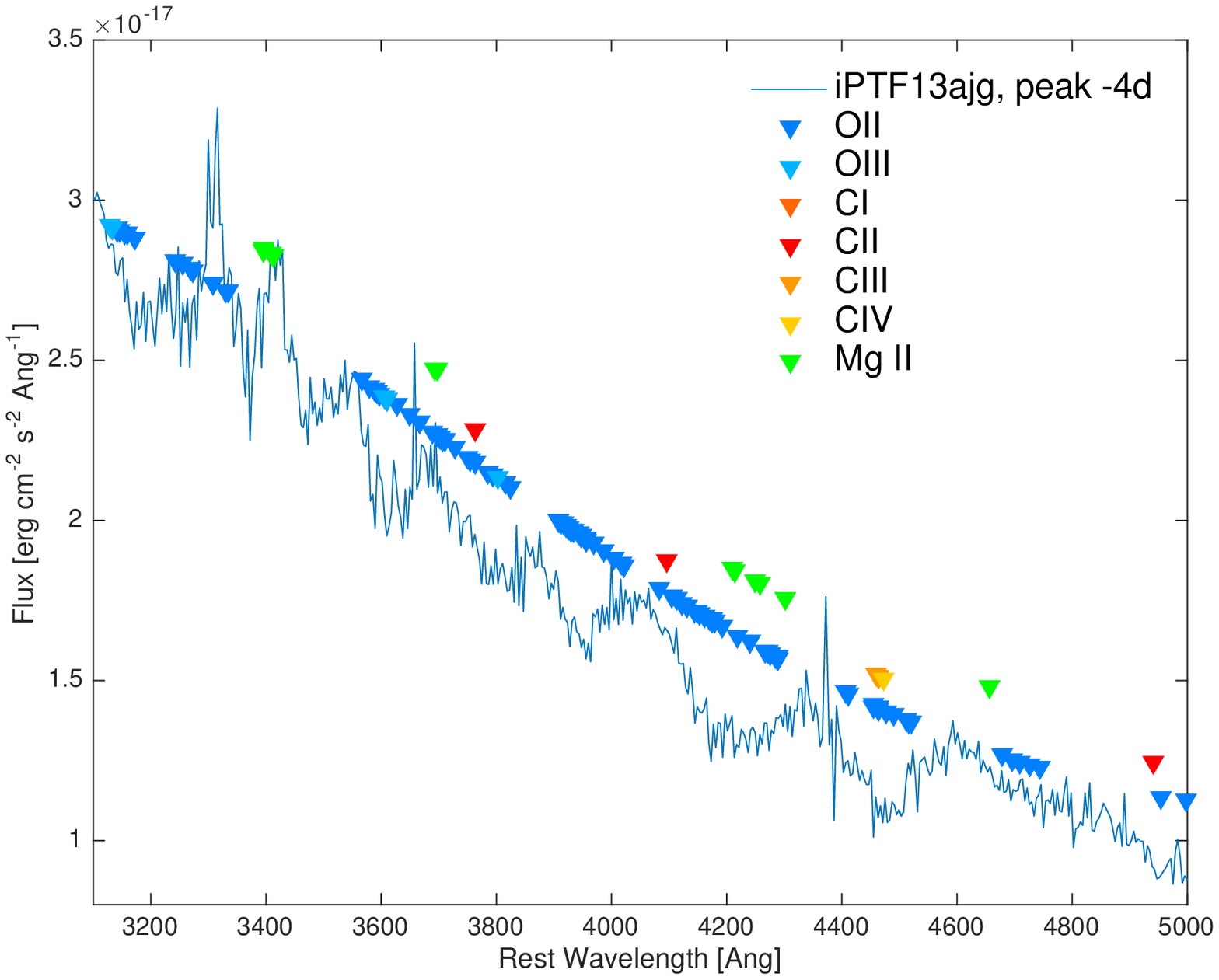}
\caption{The High SNR VLT spectrum of iPTF13ajg (from \citealt{2014ApJ...797...24V}; heavily binned to improve presentation) offers a detailed view of the restframe UV emission of SLSNe-I. Analysis of the blue portion of the visible light spectrum (right panel) indicates an expansion velocity of V$_{\rm exp}=12000$\,km\,s$^{-1}$. Inspecting the UV part of the spectrum (left) shows four broad distinct absorption troughs below $2800$\AA. In this range, O lines seem to be much less dominant. Higher ionization C lines (CIII and CIV) align well with the central two features, while Mg II and OIII may be important for the reddest and bluest features, respectively. Open symbols denote weaker CII lines (see text). }
\label{FigUV}
\end{figure*}

Due to their extreme luminosity, SLSNe-I are observable to high redshifts, making their restframe UV spectra accessible to visible-light instruments. It is therefore interesting to extend our analysis into the UV range. To do this we inspect a high SNR spectrum of the relatively high redshift (z=0.74) SLSN-I iPTF13ajg obtained by \cite{2014ApJ...797...24V}, on 2013 April 16, 4 days prior to peak. Fig.~\ref{FigUV} shows that the visible light part of the spectrum is similar to those we analyzed above and is well-explained by OII absorption; we determine an expansion velocity of V$_{\rm exp}=12000$\,km\,s$^{-1}$ by aligning the OII line distribution gaps with the observations. Inspection of the UV part (Fig.~\ref{FigUV}; left) shows four broad, distinct absorption features. OII does not seem to be important in this range. Of the ions we looked at so far, higher ionization C lines (CIII and CIV) align well with the central two features, while OIII may be important for the bluest feature. Mg II $2800$\AA\ has been proposed several times (e.g., by \citealt{2011Natur.474..487Q}, \citealt{2013ApJ...779...98H} and \citealt{2012MNRAS.426L..76D}) as the source of the reddest of the UV features. Our analysis shows that several other Mg II transitions are expected in both the UV and visible-light range, and some of these seem to be in conflict with the observed spectral features, e.g., around restframe $2550, 3400$ and $3700$\AA. 

Detailed models of this same data by \cite{2016MNRAS.458.3455M} invoke contribution by several IGEs (Ti, Fe and Co) as well as Si to fully explain these features, in addition to CII. Comparing their strong line lists to ours suggests that they find weaker CII lines (down to $35\%$ of the maximum NIST relative intensity) are important (these are below our default $50\%$ relative intensity cutoff; open symbols in Fig.~\ref{FigUV}). While our simple analysis suggests that C/O lines may suffice to explain all or most of the UV features, further work is required to determine the optimal cutoff for this type of analysis.      

\section{Analysis and Discussion}

\subsection{Dynamics}

A major surprising result from this analysis is that while the expansion velocity of SLSNe-I at peak ($\sim12000$\,km\,s$^{-1}$) is similar to that of other SNe, and in particular regular SNe Ic (e.g., \citealt{2016ApJ...827...90L}, \citealt{2017hsn..book..195G}), the velocity dispersion, manifest as the velocity width of absorption features from individual transitions, is much lower, $<1500$\,km\,s$^{-1}$ (\S~\ref{secvel}). \cite{2016ApJ...832..108M} showed that for Type Ic SNe, over a very wide range of values, the velocity dispersion is strongly correlated with the expansion velocity: events with higher blueshifted expansion velocities also tend to have much wider lines (larger velocity dispersion; their Fig. 7). We show that SLSNe-I around peak (which eventually evolve to resemble SNe Ic; e.g., \citealt{2010ApJ...724L..16P}, \citealt{2011Natur.474..487Q}), strongly break this correlation. Even compared to SN 1994I which is perhaps the most extreme ``narrow-line'' SN Ic (\citealt{2016ApJ...827...90L}), SLSNe-I at peak have much narrower lines. Analysis of peak spectra SN 1994I along the lines of Fig.~\ref{Fig_PTF12dam_red_zoom} would suggest a range of $\pm7500$\,km\,s$^{-1}$ for the $6100$\AA\ feature of SN 1994I, at least 5 times the value we measure for PTF12dam.  

This finding indicates that the emitting region in SLSNe-I around peak luminosity is very narrow in velocity space, with a velocity dispersion that is an order of magnitude less than the expansion velocity. To avoid absorption at lower velocities, material moving in relatively large angles with respect to our line of sight must not significantly absorb the photosphere. This can naturally arise from a strongly aspherical geometry (see e.g., \citealt{2016ApJ...831...79I} and \citealt{2017ApJ...837L..14L} for discussion), or else it requires an emissivity gradient in the outer envelope of the ejecta which is very different from that of other SN types.  

\subsection{Unusual events: the nature of ASASSN-15lh and SLSN-I with late-time hydrogen lines}

\begin{figure}
\hspace*{-1cm}\includegraphics[width=10.5cm]{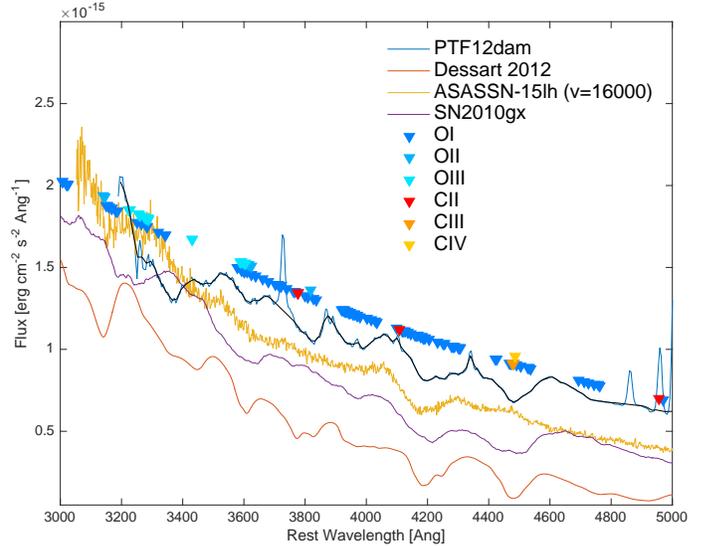}
\caption{Comparison of the early spectrum of ASASSN-15lh from \citealt{2016Sci...351..257D} with the high SNR spectrum of PTF12dam, the \citealt{2012MNRAS.426L..76D} model, and our OII line distribution. Aligning the strong feature seen in the ASASSN-15lh spectrum with the strong OII feature requires an expansion velocity of $16000$\,km\,s$^{-1}$, but the strong additional OII features seen in other objects and predicted by both our analysis and the \citealt{2012MNRAS.426L..76D} model are not seen, even though the SNR of the ASASSN-15lh spectrum is high.}
\label{Fig_15lh}
\end{figure}

The flexibility of this analysis method motivates its use on special cases of interest. One such is the recent event ASASSN-15lh. \cite{2016Sci...351..257D} presented the discovery of this very luminous transient, and suggested a classification of a SLSN-I based on spectral similarity to several events, in particular SN 2010gx (\citealt{2010ApJ...724L..16P}). We show our analysis of this early spectrum of ASASSN-15lh in Fig.~\ref{Fig_15lh}. The following is evident from this figure. First, using the well-determined redshift of ASASSN-15lh from its host spectra ($z=0.2326$), in order to align the strong feature seen at restframe $4100$\AA\ with the strong OII feature suggested by \cite{2016Sci...351..257D}, one has to assume a rather high expansion velocity for this event (V$_{\rm exp}=16000$\,km\,s$^{-1}$). 

Assuming this velocity, our analysis suggests that several additional features should be seen, both a strong additional feature to the red of the observed one, as well as additional blue features. While one can argue that the feature at restframe $4500$\AA\ has a contribution from CIII and CIV in addition to OII, and therefore that the ratio of these features might be influenced by the ejecta composition (C/O ratio), this is not the case for the bluer features according to neither our analysis nor the \cite{2012MNRAS.426L..76D} model plotted for comparison. The single strong absorption feature in the ASASSN-15lh spectrum is not consistent with OII. We note that the comparison object used by \cite{2016Sci...351..257D} (SN 2010gx, spectrum from \citealt{2018ApJ...855....2Q} shown) does show the correlation between emission peaks and gaps in the OII line distribution. We therefore conclude that our analysis of the early spectrum of ASASSN-15lh does not support its association with the spectroscopic class of SLSNe-I, and therefore perhaps favors alternative explanations (\citealt{2016NatAs...1E...2L}). 
 
\begin{figure*}
\hspace*{-1.5cm}\includegraphics[width=10.5cm]{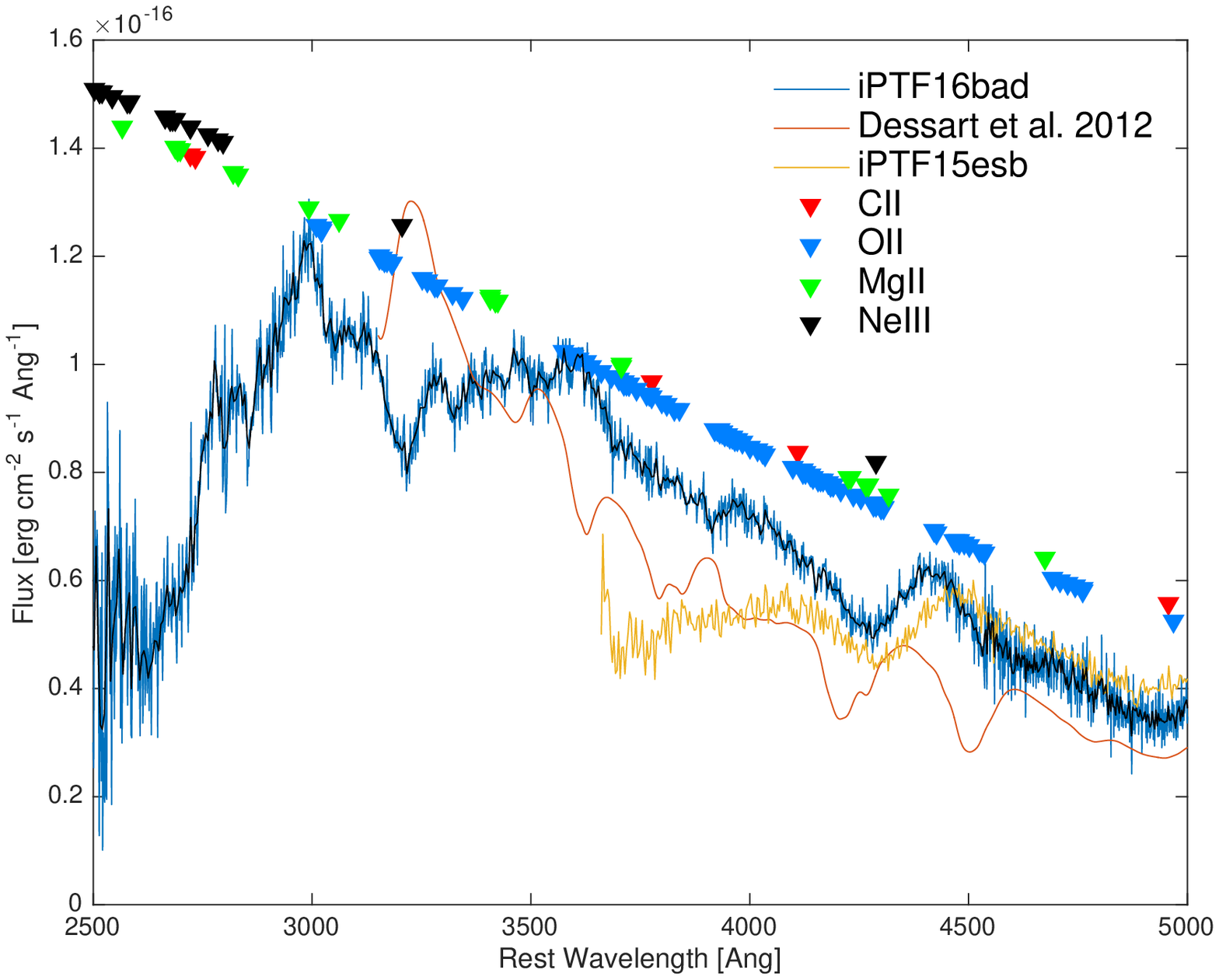}
\hspace*{-0.5cm}\includegraphics[width=10.5cm]{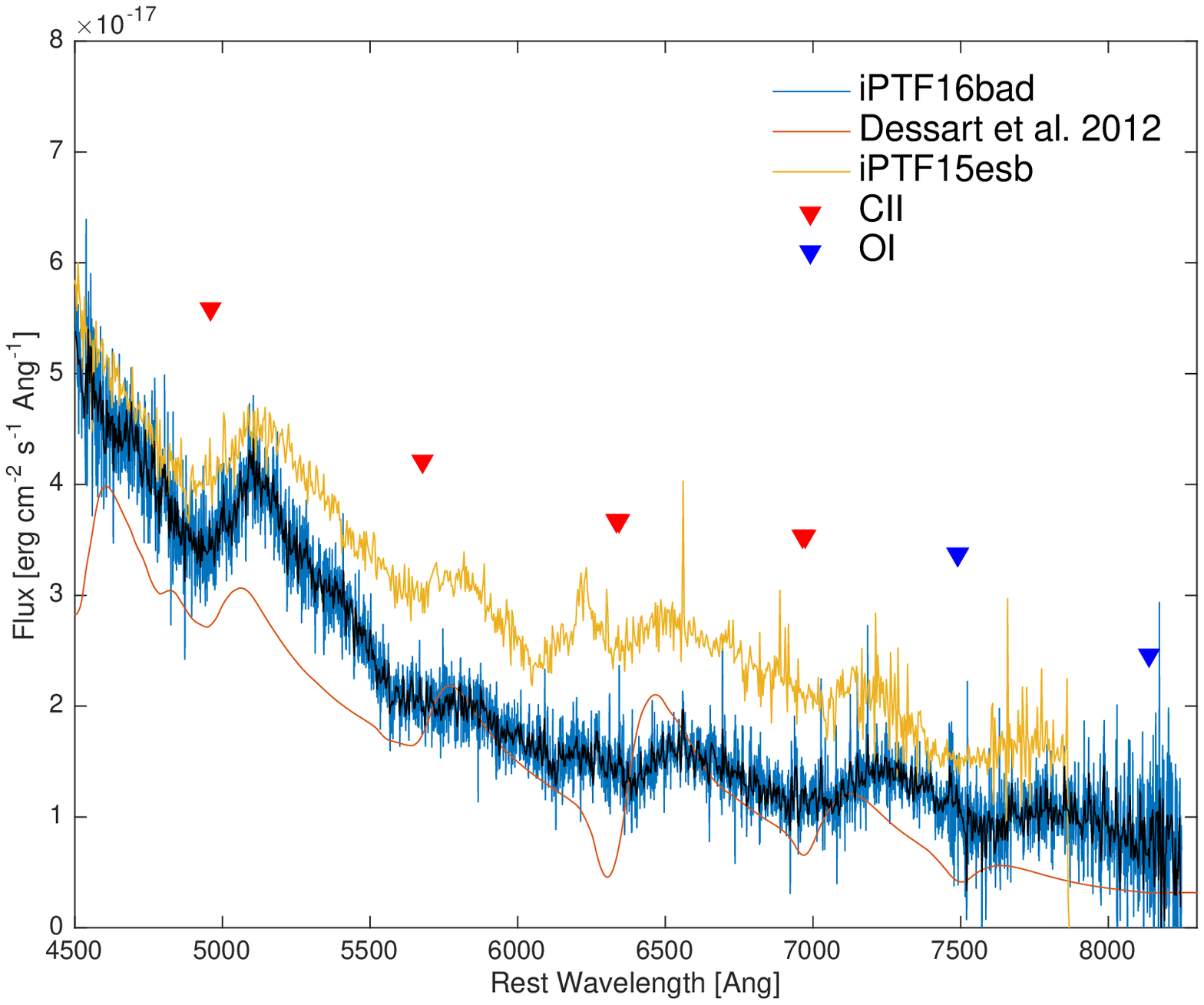}
\caption{Analysis of spectra of iPTF16bad and iPTF15esb (from \citealt{2017ApJ...848....6Y}; the iPTF15esb spectrum is binned to improve presentation). The red side (right panel) is well explained by C and O features at a single expansion velocity of $11000$\,km\,s$^{-1}$, as seen for normal SLSNe-I above. A single feature around $6000$\AA\ is seen only in the spectrum of iPTF15esb and is not consistent with C or O lines (see text). Our identification is consistent with the SLSN-I model by \cite{2012MNRAS.426L..76D}. Analysis of the blue portion of the visible light spectrum (left panel) shows significant difference with respect the SLSN-I spectra shown above. In particular, gaps in the OII absorption line distribution (light blue triangles) do not align with observed emission peaks, and features in the bluest part are also in strong conflict with the \cite{2012MNRAS.426L..76D} model. 
We illustrate an alternative model to explain most observed features by Mg II and Ne III: this is the only combination we found so far that can explain the strong features observed in the blue spectrum of iPTF16bad.}
\label{Fig16bad}
\end{figure*}

\cite{2017ApJ...848....6Y} presented several SLSNe that are similar to SLSNe-I early on, but develop strong hydrogen lines at later phases. 
In order to test the association of the objects discussed in this paper with the spectroscopic class of SLSNe-I, we apply our analysis to near-peak spectra. Figure~\ref{Fig16bad} (right) shows the red portion ($\lambda>4500$\AA) of the spectra of iPTF16bad and iPTF15esb obtained on 2016 June 7 and 2015 December 16 (2 and 7 restframe days after peak according to \citealt{2017ApJ...848....6Y}), respectively. The spectra are well explained by lines of CII and OI, at a single photospheric velocity ($11000$\,km\,s$^{-1}$). Comparison with the model of Dessart et al. (2012) validates this analysis. Interestingly, the spectrum of iPTF15esb shows an additional feature around restframe $6000$\AA\ which is missing from the spectrum of iPTF16bad. Our analysis, as well as the fact that this feature is not seen in the Dessart et al. (2012) model, suggests it is not associated with C/O. To explain this feature with hydrogen H$\alpha$ would require a very high expansion velocity ($23000$\,km\,s$^{-1}$). The fact that no absorption is seen at the expected location of H$\beta$ may argue against this interpretation.

We now turn to the blue side of the spectrum ($\lambda<4500$\AA) which for SLSNe-I is dominated by the OII absorption forest. Inspection of Fig.~\ref{Fig16bad} (left) shows that this is not the case here. The gaps in the forest of OII features do not align with the spectral features. In particular, the strong absorption seen in both iPTF16bad and iPTF15esb near resframe $4300$\AA\ is offset from the location of the strongest OII feature as seen above and expected from the Dessart et al. (2012) model. The flux peak around $4400$\AA\ does not align with a gap in the OII line distribution, but rather with a cluster of absorption lines. The same holds for the bluer features. Artificially forcing the OII gaps to align with the strongest emission peak at $4400$\AA\ requires the OII emission to arise from a photospheric region moving $5000$\,km\,s$^{-1}$ slower than the area emitting OI, which does not make sense. As an alternative to OII lines to explain the blue part of the spectra of iPTF16bad we outline a model based on Mg II and perhaps Ne III to explain the main features seen; a full investigation of this spectrum likely requires additional modelling. To conclude, while these objects show some spectral similarities to ``standard'' SLSNe-I, their peak spectra also show significant differences. In view of the evidence for recent ejection of hydrogen-rich material by the progenitors of these stars (\citealt{2015ApJ...814..108Y}, \citealt{2017ApJ...848....6Y}) the progenitors may be quite different from the pure C/O envelopes we infer for SLSNe-I based on the analysis above.

\section{Summary}

We presented the application of a simple analysis procedure designed to explain simple supernova spectra (assuming a narrow photosphere moving at a single velocity) to the peak spectra of Type I superluminous SNe (SLSNe-I). We found the following:

$\bullet$~SLSN-I near-peak visible-light spectra are well-explained by mostly C/O lines moving at a single photospheric velocity. 

$\bullet$~When available, high SNR spectra in the red portion of the visible light range allow an accurate determination of the ejecta expansion velocity via multiple individual transitions of OI and CII.

$\bullet$~Using velocities determined from the red portion of the spectra, the blue portion is well described by the effects of a thick forest of multiple OII absorption lines. Gaps in the OII line distribution align with apparent peaks in the spectra. 

$\bullet$~Alignment of the OII line distributions with observed spectral features allows to determine the ejecta velocity also for events with poorer SNR or without spectral coverage redward of restframe $5000$\AA, typical for higher-redshift SLSNe.

$\bullet$~The dominance of C/O lines in peak SLSN-I spectra suggests an emission from an almost pure C/O envelope, without significant contamination by higher-mass elements from deeper photospheric layers. 

$\bullet$~Line widths determined from spectral features associated with individual transitions suggest very low values of velocity dispersion ($<1500$\,km\,s$^{-1}$), compared to the high expansion velocities measured for these events ($>10000$\,km\,s$^{-1}$), making SLSNe-I strong outliers to the correlation between high expansion velocities and high velocity dispersions (line blueshifts and line widths) e.g., of \cite{2016ApJ...832..108M}. This indicates that the emitting photospheric layer is very narrow in velocity space.

$\bullet$~We find that the early spectrum of the peculiar superluminous event ASASSN-15lh (\citealt{2016Sci...351..257D}) is not consistent with OII absorption and is not similar in that sense to SLSNe-I, perhaps supporting an alternative explanation as a TDE (\citealt{2016NatAs...1E...2L}).  

$\bullet$~The sample of SLSNe that show hydrogen emission at late times (\citealt{2017ApJ...848....6Y}) show both similarities (redward of $5000$\AA) and significant differences (blueward of $5000$\AA) with respect to standard SLSNe-I, as may be expected from progenitors who only recently shed their outer hydrogen layers.

\acknowledgments
AGY would like to thank the hospitality of the Munich Institute for Astro- and Particle Physics (MIAPP). Much of this work was accomplished during the recent ``superluminous supernovae in the next decade" MIAPP program. This work could not have been carried out without use of the WISeREP database and
the support of O. Yaron, as well as the invaluable help of I. Manulis and N. Knezevic. 
I thank S. Schulze, G. Leloudas, A. De-Cia, S. Dong and B. Katz for useful comments on this manuscripts and
L. Yan for providing a copy of the spectrum of gaia16apd that prompted some of this work, and her
useful advice.
AGY is supported by the EU via ERC grants No. 307260 
and 725161, the Quantum Universe I-Core program by the 
Israeli Committee for Planning and Budgeting, and the ISF; 
a Binational Science Foundation "Transformative Science" grant
and by a Kimmel award.



\end{document}